# Benefits and Burdens of Technology-Supported Activity Pacing in ME/CFS: A Thematic Analysis of Reddit Discussions

Martin Matych
Department of Information Technology
Prague University of Economics and Business
Prague, Czech Republic
*martin.matych@vse.cz*

Michal Dolezel
Department of Information Technology
Prague University of Economics and Business
Prague, Czech Republic
*michal.dolezel@vse.cz*

**Technology and self-tracking play an increasingly important role in the self-management of chronic conditions. Myalgic Encephalomyelitis/Chronic Fatigue Syndrome (ME/CFS) is a type of energy-limiting condition, with activity pacing as a key self-management strategy that helps individuals regulate their activity within their energy limits to prevent symptom worsening. However, people with ME/CFS frequently adopt technologies designed for healthy bodies, which they have to repurpose to suit their pacing needs. We report findings from a thematic analysis of Reddit data, using the Non-Adoption, Abandonment, Scale-Up, Spread and Sustainability (NASSS) framework as a sensitising concept. We identify three pacing strategies among people with ME/CFS: no-tech pacing, low-tech pacing (e.g., spreadsheets and diaries), and high-tech pacing (relying on continuous passive sensing from consumer wearables), and characterise the benefits and burdens of pacing technology use. Among these, affordability was a salient topic, highlighting the importance of making technology for pacing financially accessible to people with ME/CFS who follow the high-tech route.**

## 1. INTRODUCTION

Myalgic Encephalomyelitis/Chronic Fatigue Syndrome (ME/CFS) is a debilitating chronic condition that profoundly shapes the everyday lives of people living with it (Nacul et al., 2011). The key self-management strategy for ME/CFS is activity pacing (referred to as “pacing”), which helps individuals regulate activity within their physiological limits and better manage energy use (Kos et al., 2015).

Pacing relies on awareness of activity, symptoms, and personal limits, which are often challenging to manage for people with ME/CFS themselves due to their condition (Davies et al., 2019). **Technology-supported pacing**, including self-tracking apps and wearable devices, enables people with ME/CFS to record, quantify, and visualise activity and symptom data, helping them adapt their daily routines (Rekeland et al., 2022). Beyond the physiological management, such technology can also serve as a psychological crutch, providing what Homewood, Just and Johansson (2024) describe as a "proof by numbers," offering people objective evidence of their illness to present to sceptical healthcare providers, family, or friends (Clague-Baker et al., 2023).

However, most commercially available self-tracking technologies are designed around increasing physical activity and promoting progressive workload, which does not align with the needs of people with ME/CFS (Homewood, 2023). As a result, individuals with energy-limiting conditions often repurpose these tools in unintended ways – not to optimise performance, but to understand their limits and avoid overexertion (Homewood et al., 2024).

Existing qualitative Human-Computer Interaction (HCI) research has offered valuable insights into how people with ME/CFS and long COVID, conditions sharing core symptoms, adopt, use and adapt technology-supported pacing (Clague-Baker et al., 2023; Davies et al., 2019; Homewood et al., 2024; Paymal and Homewood, 2024). Because these foundational studies primarily utilised interviews and surveys, and they naturally faced challenges regarding the fluctuating energy limits of this population (Homewood et al., 2025; Paymal and Homewood, 2024), there remains an opportunity to study how individuals use technology for pacing in their everyday lives, especially as the technology-supported pacing landscape continues to evolve (Sanal-Hayes et al., 2026).

We address this gap through a thematic analysis of Reddit discussions. Utilising this forum allows us to analyse naturally occurring, contributor-driven statements rather than data structured by a specific research protocol (Boettcher, 2021; Proferes et al., 2021), thereby building on existing literature without introducing an additional participation burden. Here, we aim for a holistic, contemporary view of technology-supported pacing through two central research questions (RQ):

- **RQ1:** What activity pacing strategies do people with ME/CFS use to learn their energy limits and self-manage their condition?
- **RQ2:** What benefits and burdens do people with ME/CFS find in technology-supported pacing, and how does this shape decisions to adopt, sustain, or abandon a digital tool?

We identify three pacing strategies among people with ME/CFS: **no-tech** pacing, **low-tech** pacing, and **high-tech** pacing, and characterise the benefits and burdens of pacing technology use.

## 2. BACKGROUND

### 2.1 Self-tracking in Chronic Illness

Self-tracking as a personal informatics practice has been extensively studied for almost two decades (Epstein et al., 2015; Li et al., 2010; Rooksby et al., 2014). While earlier work introduced stage-based, systems-oriented views on personal informatics (Li et al., 2010), later work frames self-tracking as a lived, episodic experience (Epstein et al., 2015; Rooksby et al., 2014). In HCI research focusing on chronic illness, self-tracking tends to be interpreted as a vehicle for regaining a degree of predictability and sense of control (Ayobi et al., 2017).

Self-tracking in chronic illness conceptually differs from performance-oriented self-tracking (MacLeod et al., 2013), as not all people with chronic conditions view it as a source of "joy and passion" (Ayobi et al., 2017). Also, they do not typically consider self-tracking data representing a self-improvement and self-optimisation baseline (Choe et al., 2014); instead, the data may provoke strong emotions and remind them about their illness (Ancker et al., 2015). In that sense, "finding a balance between monitoring and an excessive focus on symptoms" is a crucial coping skill (Ayobi et al., 2017), which may eventually result in a rational decision to abandon a technology (Clawson et al., 2015; Epstein et al., 2016). This is consistent with Baumer and Silberman's (2011) argument that either no technology (no tech) or a simple technology (low-tech) is a legitimate technology choice in some groups of users.

Interestingly, for high-technology users, finding the "perfect" technology for their chronic illness is a never-ending process that requires strong dedication and tech-savviness. In that sense, some people with chronic conditions tend to act as "bricoleurs," continuously exploring and adopting emerging technologies (Gibson et al., 2019). They abandon some and adopt others, and diffuse technology within their own communities – driven by their frustration with missing top-down initiatives (O'Kane et al., 2019). In that sense, they help with closing the gap between early and late adopters (Satchell and Dourish, 2009).

### 2.2 ME/CFS and Technology-Supported Pacing

ME/CFS is a complex, multisystemic condition whose incidence is expected to rise worldwide (Vester et al., 2026). The most common symptoms of ME/CFS include persistent fatigue, post-exertional malaise (PEM), unrefreshing sleep, orthostatic intolerance, and cognitive impairments such as brain fog (Chang et al., 2021). Currently, non-pharmacological self-management of ME/CFS relies on activity and rest balance planning via pacing, which encourages patients to stay within their "energy envelope" to prevent overexertion, which could otherwise lead to PEM (Jason et al., 2013; Kos et al., 2015; National Institute for Health and Care Excellence (NICE), 2021). However, subjective tracking and planning activities to avoid crashes are challenging due to their unpredictable onset, often coming only after multiple hours or even days (Chu et al., 2018).

To facilitate pacing, current clinical guidelines recommend clinicians to "make self-monitoring of activity as easy as possible by taking advantage of any tools the person already uses, such as an activity tracker, phone heart-rate monitor or diary" (National Institute for Health and Care Excellence (NICE), 2021). In practice, heart rate (HR) serves as an accessible proxy for the anaerobic threshold, allowing individuals to establish personalised boundaries to avoid triggering PEM (Clague-Baker et al., 2026). However, commercial HR monitoring lacks individualised thresholds, ignores cognitive or emotional exertion, and offers no professional guidance, highlighting a design mismatch for chronic illness (Clague-Baker et al., 2023).

Beyond core HR monitoring metrics like resting HR and heart rate variability (HRV), people with ME/CFS adopt other derived metrics to track energy limits: from simple step counts (Rekeland et al., 2022) to more ambiguous, manufacturer-defined metrics such as sleep scores or other proxies estimating energy levels and readiness (Homewood et al., 2024). However, these metrics are largely proprietary and often lack independent medical validation: a recent umbrella review found that only around 11% of commercially available wearables have been validated for even a single biometric outcome (Doherty et al., 2024), and composite scores such as sleep score are particularly opaque, as manufacturers rarely disclose their algorithms and revise them without notice (de Zambotti et al., 2019).

## 3. METHODS

Reddit is a large online platform made up of topic-specific communities called subreddits, where users start discussions with initial posts and reply to each other in threaded comments. For this study, we collected discussion threads from r/cfs subreddit, the largest Reddit community for research, treatment, and personal experience of ME/CFS, with around 90,000 weekly visitors and 6,500 weekly contributors. We searched this subreddit for posts that mentioned wearable devices or apps together with the concept of pacing, using the query "selftext:((wearable OR wearables OR device OR devices OR app OR apps) AND (pace OR pacing))". All matching posts published within a 1-year window between 1 May 2025 and 30 April 2026 were collected, resulting in 88 discussion threads. Each thread, including its full comment structure, was downloaded and converted to text using a Python script, preserving the original reply hierarchy for later analysis.

Following data collection, we applied a two-stage screening process to determine which threads and comments were included in the thematic analysis. Inclusion and exclusion criteria were developed inductively, following an initial reading of the corpus, and refined in discussion with the second author. In the first stage, we applied structural, thread-level criteria to the initial corpus of 88 threads, based on the inclusion and exclusion criteria (Appendix 1) applied to the content of the original top-level post. Screening at this stage was conducted by the first author, with edge cases discussed and resolved by consensus between the first and second authors. This stage resulted in the inclusion of 63 threads and the exclusion of 25 threads. In the second stage, comment-level filtering was conducted dynamically during the coding phase: individual comments were utilised for coding through close reading, based on their adherence to the same inclusion/exclusion criteria, their contribution to the analysis, and their relevance to the domains of the Non-adoption, Abandonment, Scale-up, Spread, and Sustainability (NASSS) framework (Greenhalgh et al., 2017). Iterative coding of 50 threads (449 posts and comments) yielded 824 code applications, at which point thematic saturation was reached as no new codes emerged.

We conducted a thematic analysis (Braun and Clarke, 2006) of the included threads and comments theoretically informed by the NASSS framework, which was used to sensitise researchers. NASSS introduces seven conceptual domains: Condition (e.g., severity, comorbidities); Technology (e.g., features, usability); Value proposition (e.g. cost and effectiveness relative to alternatives); Adopters (e.g. users' skills and preferences); Organisation (e.g. manufacturers' policy, customer support); Wider system (e.g. role of healthcare providers, family); and Embedding and adaptation over time (Greenhalgh et al., 2017). While NASSS was initially introduced by implementation science literature, it has been previously utilized also in HCI (Waddell et al., 2024).

Our qualitative coding combined deductive and inductive elements. At the top level, we used the domains of the NASSS framework (Greenhalgh et al., 2017) as a coding scaffold, since it offers a structured way to distinguish between different sources of complexity shaping technology adoption. Within each NASSS domain, codes were developed inductively, staying close to participants' own language rather than following a fixed coding scheme. To ensure consistency, the first and second authors independently coded a random sample of 10% of the included threads, then compared and discussed their codes to reach a shared codebook. The first author then coded the remaining threads using this shared codebook.

This study was considered exempt from ethical approval under the local institutional policy (PR 3/2024). Following established practice for disguising online research data (Bruckman, 2002), we further paraphrased illustrative quotations using Claude Sonnet 5 to reduce their traceability to source posts; both authors independently verified that meaning was preserved. During the manuscript writing, large language models were used for language editing and stylistic refinement.

## 4. FINDINGS

In this chapter, we present recurring themes identified through thematic analysis, organised into five thematic sections. Section 4.1 addresses RQ1, while Sections 4.2–4.5 address RQ2.

### 4.1 Learning the Limits – Three Strategies of Activity Pacing

Many contributors describe pacing as a skill that can be learned but takes years to master and demands strong discipline along the way. Several noted that the learning period itself carries risk, since improper pacing during that time can cause permanent worsening of the baseline symptoms. In the early stages, resting does not come naturally to people who were active before getting ill, as many describe a strong pull to keep doing what they used to do, and real frustration at having to hold themselves back.

> Overexertion at the start of my illness left me bedbound; I now sleep most of the day. It has been a long and hard journey.

Within this, we identified three distinct self-management strategies, which we characterise as:(i) **no-tech** pacing*,* i.e. relying on bodily sensations, (ii) **low-tech** pacing, i.e., using simple software tools such as spreadsheet tables, (iii) **high-tech** pacing*,* i.e., using apps and wearables to track activity and reveal previously unnoticed limits. Some contributors (i.) simply trust their own bodily awareness above anything a device could tell them, treating internal sensation as the most reliable pacing signal available.

> I got no benefit from pacing devices. Tuning into the body's own signals is much more effective, although recognising those subtle cues takes long practice; once learned, they are priceless.

Others (ii.) prefer keeping manual records, timing activities, or logging them in something as simple as a spreadsheet, specifically to learn their personal PEM triggers. Several who did this for a few months reported that they eventually understood the patterns well enough and no longer needed the tracking.

> For a few months I logged activities and PEM in a simple spreadsheet, which taught me a lot about my triggers; eventually the tracking became unnecessary.

For many (iii.), though, neither of these is straightforward, because PEM itself often doesn't show up until hours or even days later. Several contributors describe not recognising overexertion until it's already too late, which makes learning limits through experience alone slow and costly. This is where technology-supported pacing becomes powerful. Wearables and apps that give visual feedback, alert when a threshold has been crossed, and encourage rest before a crash comes in were described by some as transformative, changing their pacing routine entirely, helping them finally identify their triggers, and in a few cases making a real difference after decades of being unable to pace successfully.

> It [the Visible app] transformed my pacing after 30 years of illness during which I had never been able to pace successfully.

One contributor implies that technology-supported pacing tends to help two different groups of users: people new to the illness and still working out their limits, and people who already know their limits but struggle to act on that knowledge without an external impulse telling them to stop.

### 4.2 Benefits Provided by Data-driven Insights

The most prevalent theme emerging from our data concerns the value that people using technology-supported pacing derive from the data produced by these devices. For many contributors, physiological metrics such as heart rate, HRV, and derived ones like stress scores, exertion or recovery indicators turned an otherwise unpredictable condition into something actionable. Several described specific enlightening moments in which a metric revealed a hidden trigger, such as discovering that something as small as reaching for an object or standing up too quickly could spike HR well past the anaerobic threshold. This level of detail was described as impossible to notice through bodily sensation alone.

> Seeing what triggers my heart rate has been an eye-opener: small everyday movements like standing up quickly or bending over send me above my limit even during minimal activity.

A recurring pattern was using these metrics predictively, treating changes in HRV or resting HR as early warning signs of an approaching crash. This predictive value given by technologies was especially appreciated by some users, since it opened a window for preventive rest that symptom-based pacing could not offer, given the typical PEM delay.

> The app frequently anticipates my PEM before I notice anything — a morning alert about abnormal sleep metrics (elevated HR, low HRV) prompts me to take the day gently even when I feel well.

Several contributors valued having objective numbers for confirming to themselves that their exhaustion was real or for convincing sceptical doctors that their condition was physiological rather than psychological. On the other hand, frequently checking the uncompromising data has led to stress and anxiety in some people.

> Having biometric evidence that my body is genuinely struggling is validating, though the same data can also feel discouraging.

These tensions between empowerment and limitation foreshadow the persistent technical barriers described in the following section.

### 4.3 Persistent Technological Barriers

The flip side of the coin, going hand in hand with the benefits of technology, was its reported shortcomings in failing to fit the ME/CFS condition. A very frequent complaint was that HR- and HRV-based tracking captures physical exertion only, while cognitive exertion, which many contributors said triggers PEM in equally debilitating way, remains invisible to the device in most cases. Several described being told they were within energy budget according to an app while already suffering from a crash caused by mental or social effort. Some also reported that certain technologies simply were not helpful for their condition or that the metrics were too ambiguous, which undermined their trust in the tool.

> I experienced my most severe crash despite keeping within the recommended pacing points, so I didn't find them useful.

Finding the right HR threshold was itself a complicated, trial-and-error process for many, and medications such as beta blockers were reported to distort metrics even further, forcing people to manually recalibrate HR zones or rely on vendor support. A few bypassed this adopting unofficial, community-built add-ons after official app settings would not go low enough for their needs.

> Because the watch's built-in floor for a heart-rate cap (120 bpm) sat above my aerobic threshold (roughly 110), I switched to a free, regularly updated third-party pacing widget with an at-a-glance watchface.

A few contributors noted that fitness-oriented technologies actively pushed them toward overexertion, the opposite of what pacing requires, since most consumer wearables are designed for healthy or athletic users rather than an energy-limited population.

> Because Fitbit targets healthy people aiming to exercise more, its motivational prompts have to be consciously ignored.

Hardware deficiencies, such as poor wireless connection or degrading batteries, created additional barriers. In some cases, these were even caused by the unintended use of the technology towards pacing.

> Continuous activity recording is hard on the battery; running one from morning until night eventually damaged my device's battery.

Additionally, an intolerance to wearable materials was observed frequently among our data. Due to skin irritation and allergic reactions to bands or casings, some contributors were forced to switch wrist placement, try different bands, or abandon an otherwise-useful device purely for physical discomfort.

> After roughly a month of use I developed an allergic skin reaction to the sensor unit itself (not the strap), which is frustrating because the round-the-clock monitoring had really supported my pacing.

### 4.4 Weighing Price Against Perceived Value

Price tag was one of the most consistently named factors shaping technology decisions, and it interacts directly with the condition itself. Since ME/CFS frequently restricts or ends people's ability to work, many contributors described being financially dependent on partners, family, or disability income, which made any purchase or subscription, something to weigh carefully. This produced a range of affordability strategies: contributors actively sought non-subscription options (free companion apps, one-time-purchase watches, budget alternatives such as basic fitness bands), bought devices second-hand and pointed each other toward manufacturer discounts, or workarounds to reduce ongoing cost.

> The upfront cost of the band combined with a mandatory annual subscription is simply out of reach — I can't work, our household runs on a single income, and we can't gamble that much money on something unproven.

However, price was rarely judged in isolation from perceived value. Several contributors cancelled their premium subscriptions because the features they relied on, such as basic heart rate alerts, did not justify the recurring fee. This perception was strengthened when the tool failed to capture a crucial insight like cognitive exertion, making it seem poor value regardless of price. Conversely, technology explicitly designed for chronic illness or pacing, rather than repurposed fitness products, was described as more trustworthy and easier to justify financially by some contributors.

> As the sole product designed around this condition, it [Visible] has no real competitor in my view.

### 4.5 Technology Switch and Abandonment

The final theme captures how technology's perceived value shifted over time, and shows that "abandonment" rarely meant abandoning technology-supported pacing altogether. More often, contributors described switching from one device or app to another, comparing Garmin, Visible, Apple Watch, Oura, Welltory etc. against each other in detail, rather than reverting to purely manual or embodied pacing. This suggests adoption and abandonment decisions are made comparatively, against available alternatives, not simply for or against technology as such.

> Replacing Visible with a smartwatch app that alerts me instantly (sound and vibration) whenever I exceed my chosen heart-rate limit has worked much better for me.

These switches were typically driven by specific features. Contributors weighed Garmin's “body battery" against Visible's "pace points", real-time heart rate display against averaged readings, sleep-tracking accuracy, or whether a subscription model still matched their needs. The value of technology could collapse without any change in the underlying device's accuracy or in the person's own condition, purely because of a change in the interface.

> The redesign that demotes the heart-rate display in favour of pace points made the app less useful to me, and I'm not willing to keep paying a subscription for that.

A few contributors reported feeling physically or mentally better after giving up a device entirely, suggesting the tool itself, not its absence, had become the source of strain.

> Strangely, my physical wellbeing improved once I stopped wearing the armband altogether.

## 5. DISCUSSION, LIMITATIONS

Our thematic analysis proposes that people with ME/CFS typically follow three distinct routes to learning their energy limits: no-tech, low-tech, and high-tech pacing. This resonates with Baumer and Silberman's (2011) view that opting out of technology could be a reasonable choice in context of a complex condition like ME/CFS, and observation that balancing monitoring against excessive symptom focus is itself a coping skill in chronic illness (Ayobi et al., 2017). We further propose that the value of technology-supported pacing may be primarily educational rather than permanently assistive. Across the two strategies (low-tech and high-tech), digital tools helped contributors internalise their personal limits, with data complementing rather than replacing embodied knowledge. We therefore argue that abandonment following successful learning should be viewed as a positive outcome, not a failure of adoption.

In comparison with previous related literature, we found that seven years after Davies et al. (2019), people with ME/CFS still appropriate commercial fitness technologies against their intended purpose, using them to constrain rather than increase activity. Consistent with Homewood et al. (2024), mainstream wearables (Garmin, Fitbit, Apple Watch) remain prevalent despite this mismatch. However, our data show the ecosystem is diversifying toward new functionalities, emerging third-party apps, community-built add-ons and condition-specific tools such as Visible, where people with ME/CFS navigate as resourceful "bricoleurs" (Gibson et al., 2019).

The value contributors placed on physiological data extends work on self-tracking as a route to predictability and control in chronic illness (Ayobi et al., 2017), diverging from performance-oriented quantified-self framings (Choe et al., 2014; MacLeod et al., 2013). Our contributors tracked to avoid exertion, not to optimise it. Data insights, however, served a double-edged role, being both validating and distressing at the same time (Ancker et al., 2015). Persistent technological gaps mirror earlier critiques of the poor fit of consumer HR monitoring metrics for ME/CFS (Clague-Baker et al., 2023) and of manufacturers' unvalidated, ambiguous algorithms (de Zambotti et al., 2019; Doherty et al., 2024). Also, the inability of wearable devices to capture cognitive and social exertion persists. Although wearable assessment methods of cognitive load exist, they remain understudied (Zhang et al., 2026). We further identify hardware biocompatibility issues such as allergic skin reactions to wearables, which is a known phenomenon (Hussain et al., 2025), but rarely discussed in HCI.

The cost of technology interacts directly with the condition, as the ME/CFS makes it difficult for people to earn money. We found that the technology adoption, switching or abandonment is often driven by weighing purchase price against perceived value (Lazar et al., 2015). Building on Paymal and Homewood's (2024) account of pacing as daily decision-making, we show that long-term adoption and abandonment also appear as an ongoing, comparative process of experimentation. Moreover, we observed that switching or dropping a digital tool rarely meant reverting to no technology in ME/CFS.

However, this study has several limitations. Our data come from a self-selected, English-speaking subset of Reddit users with unverified, often self-reported ME/CFS diagnoses, which mostly cannot be stratified by severity, despite severity clearly shaping technology access and use in our data. As a result, the sample is likely more digitally literate and proactive in managing health than the broader ME/CFS population. At the same time, this may offer useful insights, as the workarounds developed by these early adopters, as well as their reasons for abandoning technology, can highlight barriers that may also affect wider adoption. In addition, people who stopped using technology for pacing and disengaged from r/cfs are likely underrepresented compared with those who continued to share their experiences.

## 6. ACKNOWLEDGEMENT

This research was supported by the Internal Grant Agency of the Prague University of Economics and Business under project number 44/2025.

**Appendix 1.** Inclusion (I) and Exclusion (E) Criteria

| | **Inclusion** | | **Exclusion** |
|---|---|---|---|
| *I1* | *personal experience technology use for pacing* | *E1* | *technology use unrelated to activity pacing* |
| *I2* | *a request for technology recommendations related to pacing* | *E2* | *primary focus on clinical or pharmacological interventions* |
| *I3* | *expectations or needs regarding technology design* | *E3* | *purely non-technical self-management of symptoms* |
| *I4* | *socio-technical barriers to pacing* | *E4* | *commercial promotion or other spam* |